\documentclass[pra,twocolumn,showpacs]{revtex4}
\usepackage{graphicx}
\usepackage{dcolumn}
\usepackage{amsmath}
\usepackage{bm}
\begin{document}

\preprint{cond-mat/010000}

\title{Wave mixing of hybrid Bogoliubov modes
in a Bose-Einstein condensate}

\author{Dermot McPeake$^{\dag *}$ }
\author{J F McCann $^{*}$}
\affiliation{$^{*}$ Department of Applied Mathematics and Theoretical
Physics, Queen's University Belfast, Belfast BT7 1NN, 
Northern Ireland \\
$^{\dag}$ NMRC, University College Cork,
Lee Maltings,
Prospect Row,
Cork,
Ireland.}

\date{October 2002}.

\begin{abstract}
Mode-mixing  of coherent excitations of a 
trapped Bose-Einstein condensate is modelled 
using the Bogoliubov approximation.
Hybridization of the modes of the breather ($l=0$)
and surface ($l=4$) states leads to the formation of a 
Bogoliubov dark state.
Calculations are presented for second-harmonic
generation between the two lowest-lying even-parity $m=0$
modes in an oblate spheroidal trap.
Two hybrid modes are strongly excited near second-harmonic resonance,
 and the  coupling strength, 
and hence conversion rate, to the breather mode is  half that given by an 
equivalent hydrodynamic estimate.

\end{abstract}


\maketitle

Observations of  coherence of matter wave fields
such as four-wave 
mixing \cite{4wm}, squeezing \cite{kasevich} and 
harmonic generation  \cite{oxfexp}, in atomic Bose-Einstein condensates
have recently been reported. 
The well-defined phase of the condensate means that 
small amplitude quasiparticle excitations can be produced 
that combine, through elastic coherent (phase preserving) atomic collisions, to produce 
frequency-mixed modes. In this Letter we analyse,
within the Bogoliubov model \cite{nonlin_mix},
second-harmonic generation of excitations in trapped 
spheroidal atomic  condensates: a   
process recently observed in experiment \cite{oxfexp}. 
We find  the quantal results 
are significantly  different from
hydrodynamic predictions \cite{teo_rev}. In addition to 
the quantal frequency shifts, we observe  mode hybridization 
due to coupling of degenerate Bogoliubov states and the 
creation of coherent-matter dark states. The 
hybridization process strongly affects  
second-harmonic conversion. 
Two modes are strongly excited near resonance, and the  coupling strength, 
and hence conversion rate, to the breather mode is  half that given by an 
equivalent hydrodynamic estimate.

Consider then a finite number of atoms, $N$, each of mass $m_a$, occupying a 
Bose condensate trapped by a spheroidal 
potential such that the angular 
frequency along the polar $z-$axis, axis of cylindrical symmetry,
is $\omega_z$, and the corresponding radial frequency is 
$\omega_r$.  We define the trap aspect ratio as $\lambda 
\equiv \omega_z / \omega_r$, so that the trapping 
potential can be written: $V_{\rm trap}= {\textstyle {1 \over 2}} m_a \omega_r^2 
(r^2 + \lambda^2 z^2)$.
The symmetry means that, in the limit of small
amplitude excitations,  
the acoustic equation is separable and
the  axial, $z$, radial $r$ and angular coordinate $\theta$ 
provide good hydrodynamic quantum numbers \cite{hutch,ballagh}. The 
azimuthal angular momentum quantum number is denoted by $m$.

The lowest-frequency  excitations of symmetry $m=0$, are the
quadrupole ($\omega_1$) and breathing ($\omega_2$) modes, and in the
$N$-independent hydrodynamic approximation \cite{dalfovo97}: 
$\omega_{1,2}^2 ={\textstyle {1 \over 2}} \omega_r^2 (4+3\lambda^2 \pm 
\sqrt{9\lambda^4 -16\lambda^2+16})$. The
nonlinear interactions caused by collisions create mode  
 mixing. This coupling is most strongly  pronounced when 
temporal phase-matching occurs, 
that is when second-harmonic resonance arises: $\omega_2=2\omega_1$. 
The corresponding values of $\lambda$, the trap aspect ratio,  are 
thus  $\lambda_r={\textstyle {1 \over 6}}\sqrt{  77 \pm 5 \sqrt{145}}\approx
0.683$ and $1.952$ \cite{dalfovo97}. 

The effect was first observed in an oblate trap by 
Foot and coworkers \cite{oxfexp} 
for the following parameters:
$N \sim 20000$ atoms of $^{87}$Rb,
$T < 0.5 T_c$, with  
$\omega_r=   2\pi \times 126 {\rm Hz}$. In measurements 
carried out within the range  $1.6 < \lambda < 2.8$, 
strongly-enhanced 
second-harmonic generation was found when $\lambda=1.93\pm 0.02$. 
The corresponding chemical potential of this condensate $\mu$, 
 such that $ \mu \gg \hbar \omega_r$,  lie within the range 
where hydrodynamic theory should be valid. And indeed both  the 
frequency of the breathing mode, $\omega_2$, \cite{hutch} and 
the resonant aspect ratio, $\lambda_r$, 
measured by experiment are in very good agreement 
with linear hydrodynamic theory \cite{dalfovo97}. 
Subsequently  a hydrodynamic model of the
nonlinear mixing of quasiparticles 
\cite{nonlin_mix} was  applied to 
estimate the rate of second-harmonic generation
\cite{oxfthy} and gave results consistent with experiment.
 However, this consistency hides potentially important quantal features not 
previously considered or observed.
This Letter examines the process of mode mixing and harmonic 
generation using a detailed quantal 
treatment of the process \cite{nonlin_mix}  taking 
into account atom number and quantum pressure corrections. 

\begin{figure}
\centering
\includegraphics[width=8cm]{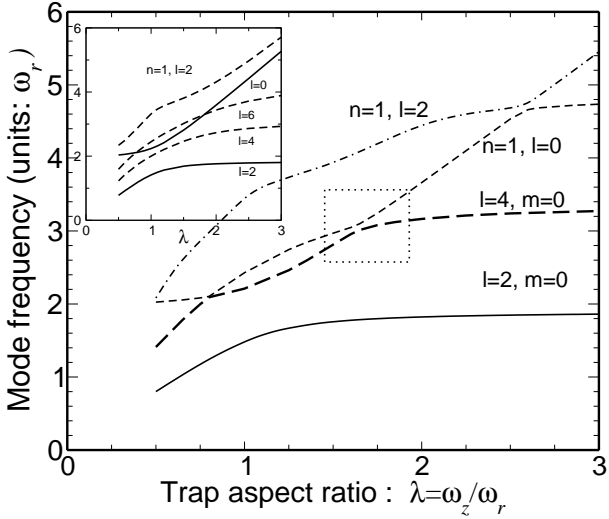}
\caption{Frequency spectrum of the 
the lowest four even-parity,  $m$=0,  excitations. 
The frequencies are in units of the 
radial frequency, $\omega_{r}$.
Bogoliubov theory 
for $N=4650$ atoms, $C=1000$, is compared with  hydrodynamic
predictions (inset) ($C\rightarrow \infty$). The
 labelling of the curves in the main graph 
corresponds to the quantum numbers for 
a spherical trap: $\lambda=1$. 
The frequency of the ($+$) hybrid mode is marked 
by the short-dashed line, and the ($-$) hybrid mode by the long-dashed line. 
The lowest frequency is the $l=2,m=0$ quadrupole (solid line).
Hybridization of the Bogoliubov breather mode $l=0$ occurs near 
$\lambda=0.80, 1.63, 2.60$. Second-harmonic resonance 
between the quadrupole and ($+$) hybrid occurs at $\lambda=1.989$.
}
\label{fig:ecurve}
\end{figure}

\begin{figure}
\centering
\includegraphics[width=8cm]{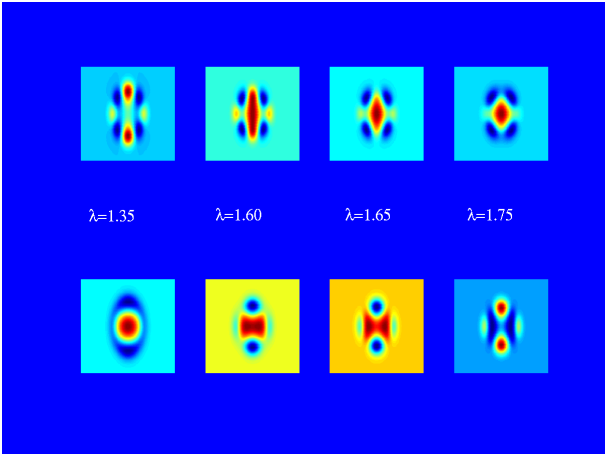}
\caption{Hybridization and formation of the Bogoliubov dark state.
The normalized  quasiparticle amplitudes $u(r,z)$ are plotted 
as a function of radial coordinate $r$ (horizontal axis) 
and axial coordinate $z$ (vertical axis) for $C=1000$.
Red  indicates positive values and  blue  negative values. 
Values of $\lambda$ increase from left to right:
$\lambda=1.35,1.60,1.65,1.75$. Below the crossing, 
at $\lambda$=1.35, the surface (upper) and breather (lower) modes are distinct. 
At  $\lambda = 1.65 $ the modes combine to form 
the dark state ($+$) hybrid. As 
$\lambda$ increases the monopole symmetry transfers
to the upper($+$) hybrid  (Fig. \ref{fig:ecurve})}
\label{fig:uplots}
\end{figure}

The formalism for wave-mixing processes in Bose condensed gases
 was developed by Burnett and coworkers \cite{nonlin_mix}.  
At low temperatures, $T/T_c \leq 0.5 $, the dynamics of Bose condensed 
gases are dominated by single quasi-particle excitations. The atom 
interactions are represented by an $s-$wave pseudopotential corresponding 
to a scattering length, $a_s$. The spectrum 
and mode densities of collective excitations 
can be obtained from the Hartree variational principle:
\begin{equation}
\delta \int dt \left(\psi, [ H_0 +
\textstyle{1 \over 2} g \psi^* \psi-i\hbar\partial_t ]\psi \right) =0
\label{vp}
\end{equation}
where $g=(4\pi \hbar^2 / m_a) N a_s$, 
$H_0 = -(\hbar^2/2m_a)\nabla^2+V_{\rm trap}-\mu$, and 
$\mu$ plays the role of a Lagrange multiplier.
The condensate and excited modes 
can be described by the linear response ansatz \cite{nonlin_mix} where:
\begin{eqnarray}
\psi(\bm{r},t) & =  & b_0(t) \phi ({\bm r}) \\ \nonumber
  &+&  \sum_{j>0} \left[ b_j(t)  {u}_j(\bm{r}) e^{-i\omega_j t}
+ b^*_j(t) {v}^*_j (\bm{r}) e^{+i\omega_jt}\right] 
\label{ansatz}
\end{eqnarray}
and where $\phi$ represents the highly-occupied condensate; that is,
$ |b_0| \approx \sqrt{N} \gg |b_j|$, $j>0$.
From the variation $\delta \phi^*$, and 
linear expansion  in the small parameters $b_j,b_j^*$ taken as 
constant, the stationary Gross-Pitaevskii equation and 
Bogoliubov equations follow:
\begin{equation}
H_0 \phi +g |\phi|^2\phi =0
\label{gpe}
\end{equation}
with $ (\phi, \phi) =1$. 
The Bogoliubov modes are solutions of the
coupled linear equations:
\begin{eqnarray}
(H_0+2g |\phi|^2) u_j +g \phi^2\  v_j & = & +\hbar\omega_j u_j 
\label{bdg1} \\
(H_0+2g |\phi|^2)v_j +g \phi^{*2} u_j & = & -\hbar\omega_j v_j 
\label{bdg2}
\end{eqnarray}
We take the conventional normalization of modes:
$(u_i,u_j)-(v_i,v_j) =\delta_{ij}$. Although the spectrum $\omega_j$ 
is unique, the mode amplitudes may contain arbitrary components of the 
kernel, the $\omega=0$  component. After 
diagonalization, it is convenient, though not necessary, to enforce orthogonality \cite{nonlin_mix}. 
Gram-Schmidt orthogonalization 
ensures the eigenmodes do not overlap the condensate:
$ \tilde{u}_j =u_j -(\phi,u_j)\phi$ and 
$ \tilde{v}_j =v_j +(\phi,u_j)\phi^*$. Suppose that 
the condensate (mode 0) contains $N$ particles, and a single 
quasiparticle mode  $i$ of excitation 
is weakly populated. Then mode-mixing collisions of the type, $i+i \rightarrow 0 +j$, 
and all crossing symmetries, will be significant in redistribution
of population if phases (energies) match and/or the scattering 
amplitude is large. 

Allowing for variation of the constants $b_j$ of the 
trial function (\ref{ansatz}) 
in the variational principle (\ref{vp}), and 
neglecting transitions far from resonance,  then we get \cite{nonlin_mix}
\begin{eqnarray}
i\hbar\frac{db_{j}}{dt} &=& \sum_{j} g M_{ij}
b_{i}^{\ast}b_{j}e^{-i\Delta_{ij}t} 
\label{motion1}\\
i\hbar\frac{db_{j}}{dt} &=& {\textstyle{ 1 \over 2}}
g M_{ij}^{\ast} b_{i}^{2} e^{i\Delta_{ij}t} 
\label{motion2}
\end{eqnarray}
where $\Delta_{ij}= \omega_{j}-2\omega_{i}$ is the detuning.  
The mode conversion equations are exactly analogous to 
those in classical nonlinear optics,  allowing for 
 two or more excited modes ($j$) to be populated from the pump mode ($i$). 
The coupling strength (scattering amplitude) for 
the process, $M_{ij}$, is given by \cite{nonlin_mix}:
\begin{eqnarray}
M_{ij} = &2\ \int d\bm{r}\{\phi^{\ast}
[2\tilde{u}_{i} ^\ast\tilde{v}_{i} ^\ast\tilde{u}_{j}
 + \tilde{v}_{i} ^\ast\tilde{v}_{i} ^\ast \tilde{v}_{j}] \nonumber \\
&+\phi[2\tilde{u}_{i} ^\ast\tilde{v}_{i} ^\ast\tilde{v}_{j} + 
\tilde{u}_{i} ^\ast\tilde{u}_{i} ^\ast\tilde{u}_{j} ]\}
\label{mij}
\end{eqnarray}
and for a condensate with uniform phase throughout, 
all  modes can be written in term of  real functions. 

\begin{figure}
\centering
\includegraphics[width=7cm,angle=0]{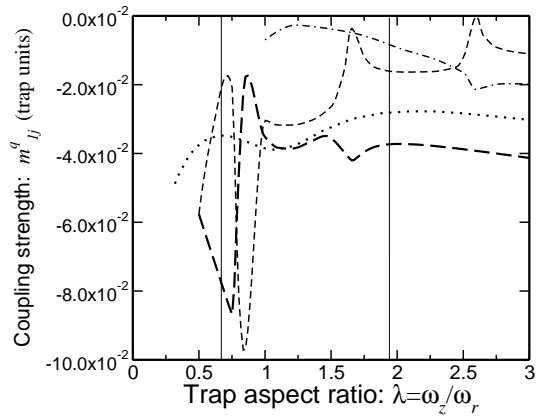}
\caption{ Coupling strength $m_{ij}$ from the 
$m=0$ quadrupole state for $C=1000$ as a function of 
trap aspect ratio $\lambda$:
Coupling to ($+$)  hybrid mode (short-dashed line),
to the ($-$) hybrid mode (long-dashed line), and to 
the $n=1,l=2$ state (dot-dashed line). The 
hydrodynamic calculation for the breather mode  \cite{oxfthy} 
is shown as the dotted line. 
Second-harmonic resonance for the ($+$)  hybrid mode  occurs 
near $\lambda$=0.68 and $\lambda$=1.95 according to 
hydrodynamic theory \cite{dalfovo97} and fig.(\ref{fig:ecurve}) as
indicated by the vertical lines.
}
\label{fig:coup}
\end{figure}

The quasiparticle amplitudes and excitation spectra were 
found through discretization of the set of equations (\ref{gpe},\ref{bdg1},\ref{bdg2})
 using the method of \cite{gyro}.
We define a dimensionless interaction parameter, proportional to 
the number of atoms condensed: 
$ C \equiv  8\pi N a_s  (\hbar/2m_a\omega_r)^{-{1 \over
2}}$, so that $C \rightarrow 0$ represents the ideal gas limit.
In the hydrodynamic approximation, 
$\mu^h = (15C\lambda/64\pi)^{2/5}\hbar \omega_r $, and the hydrodynamic 
r\'egime can be very roughly characterised by 
$\chi \equiv (\mu^h/\hbar\omega_r) \gg 1$.
An example of the spectra of Bogoliubov states 
for $C=1000$ ( that is $\chi \sim 7$)  is shown 
in fig.\ref{fig:ecurve}. 
This would  correspond, for example,  to 
 $N \sim 4500 $ atoms of $^{87}$Rb with 
scattering length $a\sim 110 a_{0}$ within a 
trap of frequency $\omega_{r}=2\pi\times 126$Hz. 
Inset, for reference, are the equivalent $N$-independent 
hydrodynamic results. The main feature of the quantal 
spectrum, compared with the hydrodynamic results, are the 
very large differences in the high-$l$ 
state frequencies. We find that in the 
quantal r\'egime these  surface (high-$l$) modes, in 
particular the state labeled $l=4$, interact 
with the second-harmonic breather mode and play an important 
role in harmonic generation.

For low-$n,l$ states, the quantum  corrections for 
the frequencies are rather small for this number of atoms. 
For example, at $\lambda=1$
the  solution of equations (\ref{bdg1},\ref{bdg2}) for the quadrupole state  
 gives $\omega_1= 1.4808 \omega_{r}$ for $C=1000$. This 
changes to  $\omega_1=1.457 \omega_r$ for $C=2000$, and reaches the 
hydrodynamic limit ($C\rightarrow \infty$) at $\omega_1= 1.4142 \omega_r$.
The $n=1,l=0$ breather frequency 
is  $\omega_2=2.2101\omega_{r}$ for $C=1000$,  shifting to 
$\omega_2=2.2195$ for $C=2000$, tending to  $\omega_2= 2.2361\omega_r$ for 
$C \rightarrow \infty$. However while the quantal 
frequencies for these low-lying modes agree well  with hydrodynamic theory  for $C \sim 1000$, 
the agreement does not extend to wave-mixing process where 
quantal effects are much more pronounced.

The  quantal breather state undergoes a series of avoided crossings with
higher-$l$ states as $\lambda$ varies (figure \ref{fig:ecurve}). 
The frequency of the $l=4$ quantal modes is significantly different from   
the hydrodynamic predictions for this r\'egime. 
For example, at $\lambda=1$  the $ l=4,m=0$ mode has 
angular  frequency
$\omega=  2.426\omega_r $ ($C=1000$),
and $\omega=  2.291 \omega_r $ ($C=2000$) compared with 
$\omega=  2.000\omega_r $ as $C\rightarrow \infty$.
While at $\lambda=2$,
$\omega =  3.164 \omega_r $ ($C=1000$), and 
$\omega = 3.042 \omega_r $ ($C=2000$), compared with 
$\omega =  2.732\omega_r $ as $C\rightarrow \infty$.
As a result, hydrodynamic theory does not predict a
 degeneracy of the $l=4$ and monopole states.

\begin{figure}
\centering
\includegraphics[width=5cm,angle=0]{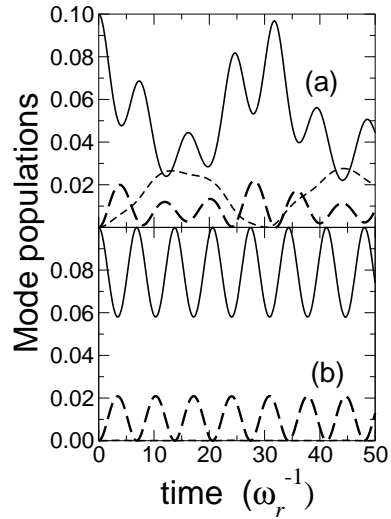}
\caption{ 
Mode conversion for $C=1000$ and 10\% pump population:
(a)  $\lambda = 2.00$: 
Population of fundamental ($n=0,l=2,m=0$) quadrupole mode (full line).
Population of  the off-resonant ($-$)  mode ($n=0,l=4,m=0$) (long-dashed line).
Population of  resonant second-harmonic ($+$)  mode $n=1, l=0, m=0$ (short-dashed line)\\
(b) Mode conversion for $\lambda = 1.65$. 
 The ($+$)   dark-state hybrid  has negligible conversion:
Population of fundamental (quadrupole) mode (full line). 
Population of ($+$)   dark-state hybrid  (short-dashed line).
Population of ($-$)   hybrid (long-dashed line). 
}
\label{fig:conversion}
\end{figure}

Near the avoided crossings in figure \ref{fig:ecurve}, the Bogoliubov 
modes mix symmetries. The  hybridization of the modes is illustrated by the 
the excitation functions, $u(r,z)$, shown in
fig.\ref{fig:uplots}; a key finding of this Letter.
Below the crossing, at $\lambda=1.35$, the $l=0$ and $l=4$ 
amplitudes resemble the spherical pattern
and the $l$-labelling is certainly appropriate. 
The radial density of the $l=4$ mode is dominated by the presence
of the centrifugal barrier pushing the mean radius
towards the surface of the condensate. At $\lambda=1.60$ (fig. \ref{fig:uplots})
the $l=0$ and $l=4$ states hybridize and lose the character of 
conventional classification schemes \cite{hutch,ballagh}.
Above the degeneracy the modes separate 
and regain their character. Identical features were found at the other crossings near
$\lambda=0.80$ and $\lambda= 2.60$ corresponding 
to interaction with $l=4$ and   $n=1,l=2$ states, respectively.
The hybridization, although 
primarily a quantal feature, persists for much larger numbers of atoms. 
Considering $C=2000$, that is $N\sim 10^{4}$, the second-harmonic 
resonance occurs at $\lambda_r=1.980$ and we find a slight displacement 
of the avoided crossing from $\lambda_c=1.65$ to $\lambda=1.54$, but 
hybridization of the type discussed is still present.

The results for the coupling strengths are shown in figure \ref{fig:coup}. The 
data are presented in scaled dimensionless units  in order to 
compare with hydrodynamic theory \cite{oxfthy}:
$m_{ij}=2 (\hbar/m_a\omega_r)^{3/2} (\mu^h/\hbar \omega_r) \lambda^{-5/6}M_{ij}$.   
The states are labelled and identified by the adiabatic noncrossing curves  in figure 
\ref{fig:ecurve}. At the degeneracy $\lambda \sim 1.65$  the strong mixing of states
is reflected in the changes in  coupling strength. 
Coupling to the upper ($+$) hybrid mode drops suddenly at this point
due to destructive interference between the mode 
components.  The effect is analogous to an optically 
dark state, that is a coherent superposition 
of states such that the dipole moments cancel \cite{scullybook}.
The interference effect is still apparent near $\lambda=2$ 
when the state is predominantly breather-like and resonant with the 
quadrupole second harmonic. In contrast the off-resonant ($-$) hybrid, 
which is surface-like for $\lambda \sim 2$ is strongly coupled 
near the degeneracy. 
 
Similar effects, including the formation of another dark state, occur 
at $\lambda \approx 2.6$ corresponding 
to the crossing between $l=0$ and $l=6$ as shown in 
figure \ref{fig:coup}.  The result of 
hydrodynamic theory  for the quadrupole to breather 
coupling strength \cite{oxfthy} is shown in  figure \ref{fig:coup} 
as the dotted line. The agreement with the quantal calculation is quite good 
near $\lambda=1$ but near and beyond the crossing point $\lambda_c=1.65$, 
the hydrodynamic model does not reflect the rapid 
quantal variation due to hybridization. At the
harmonic generation resonance near $\lambda=2$ the hydrodynamic 
model severely overestimates the coupling strength of the breather 
mode by a factor of two, and neglects the contribution of the $l=4$ state.

The solutions of the  equations (\ref{motion1},\ref{motion2}) describing
the relative populations in modes $i$ and $j$ are well known
\cite{armstrong}. For a resonant two-state model 
the characteristic coupling time  is
$T=\left|\sqrt{2}\hbar/g M_{ij}b^{r}_{i}(0)\right|$. 
In terms of the dimensionless matrix 
 elements $m^{q}_{ij}$ this can be written;
\begin{equation}
T=\left|\frac{15\lambda^{-\frac{13}{30}}}{8\pi m^{q}_{ij}b^{r}_{i}(0)
\omega_{r}}\left(\frac{15C}{64\pi}\right)^{-3/5} \right| 
\end{equation}
For $\lambda \approx 1.95$, and $C=1000$ the quantal 
prediction is $|m^{q}_{12}|$=0.013 compared with 
the hydrodynamic  result \cite{oxfthy} $|m^{h}_{12}|$=0.028 
(the dotted  line in figure \ref{fig:coup}).
Suppose that only 5\% of atoms were initially in the pump mode, then $|b^{r}_{1}(0)|^{2}$=0.05.
It follows, for $C=1000$ and $\lambda=1.95$, that the respective coupling times are 
$T^q$=14.60 ms compared with  $T^{h}$=6.78 ms, which is a substantial 
and measurable difference.
However two excited hybrid modes are present near second-harmonic 
resonance. A numerical solution of equations (\ref{motion1},\ref{motion2}) for a 
pump population of 10\% at $\lambda=2$ shows the interplay 
of the hybrid modes: figure  \ref{fig:conversion} (a). 
The nonresonant surface hybrid grows faster due to
the larger coupling strength, but is not as efficiently  
converted as the resonant  
breather hybrid state. The fundamental mode revives \cite{nonlin_mix} 
after $\sim 40$ms for these parameters. Since the two 
hybrid modes differ in frequency and symmetry 
this phenomenon  should be detectable by observation of the radial and axial 
density variations as developed in current 
experiments \cite{oxfexp}. In  figure  \ref{fig:conversion} (b) corresponding 
to $\lambda=1.65$, very low conversion efficiency is observed due to poor phase-matching. 
Moreover, in this case the dark-state hybrid  (short-dashed line) is completely suppressed. 

In conclusion, we find the hybridization of Bogoliubov modes 
plays a vital role in mode mixing, both in the coupling 
strength and resonant frequency. Another consequence is the creation of dark coherent states.
 The conversion rate for second-harmonic generation to 
the breather mode is substantially lower than the 
hydrodynamic theory, while the  off-resonant surface 
mode is converted with almost equal efficiency. 
This hybridization persists for  much larger numbers 
($N\sim 10^{4}$) of atoms   where quantal  effects were thought to be neglible.
Another potentially interesting case \cite{graham98} is 
 second-harmonic resonant coupling between the $m=2$ quadrupole and 
the $m=0$ monopole mode. In this case  an
accidental degeneracy of the monopole with a $m=4$ mode 
occurs exactly
at the predicted resonance  $\lambda=\sqrt{16/7}$ and 
the effects of hybridization would be more extreme. The
investigation of this particular transition would provide 
the clearest experimental evidence of the effect 
we have discussed.


\bibliography{mode_prl}
\end{document}